\documentclass[12pt]{article}
\usepackage{epsf}
\def\beq{\begin{equation}}
\def\eeq{\end{equation}}

\def\jp{J/\psi}

\begin{document}
\begin{titlepage}
\begin{center}
{\Large \bf William I. Fine Theoretical Physics Institute \\
University of Minnesota \\}
\end{center}
\vspace{0.2in}
\begin{flushright}
FTPI-MINN-06/33-T \\
UMN-TH-2520-06 \\
September 2006 \\
\end{flushright}
\vspace{0.3in}
\begin{center}
{\Large \bf $e^+e^- \to \gamma X(3872)$ near the $D^* {\bar D}^*$ threshold
\\}
\vspace{0.2in}
{\bf S. Dubynskiy \\}
School of Physics and Astronomy, University of Minnesota, \\ Minneapolis, MN
55455 \\
and \\
{\bf M.B. Voloshin  \\ }
William I. Fine Theoretical Physics Institute, University of
Minnesota,\\ Minneapolis, MN 55455 \\
and \\
Institute of Theoretical and Experimental Physics, Moscow, 117218
\\[0.2in]
\end{center}

\begin{abstract}
We evaluate the cross section of the process $e^+e^- \to \gamma X(3872)$ in
terms of the content of the $D^{*0} {\bar D}^0$ `molecular' component in the
wave function of the resonance $X(3872)$. If this component is dominating, the
cross section of the reaction $e^+e^- \to \gamma X(3872)$ can reach up to about
$10^{-3}$ of that for $e^+e^- \to D^* {\bar D}^*$ at energy slightly above the
$D^* {\bar D}^*$ threshold, and the considered process can be a realistic source
of the $X(3872)$ particles for the studies of this resonance.
\end{abstract}

\end{titlepage}

The extreme proximity of the mass of the $X(3872)$ resonance\cite{pdg} to the
$D^0 {\bar D}^{*0}$ threshold, as well as the co-existence of the decays
$X(3872) \to \pi^+ \pi^- \jp\,$ and  $X(3872) \to \pi^+ \pi^- \pi^0
\jp\,$\cite{belle}, strongly suggests\cite{cp,ps,mv,nat2} a significant presence
of a  `molecular'\cite{ov} $D^0 {\bar D}^{*0} +
D^{*0} {\bar D}^0$ component in the wave function of $X(3872)$. Such unusual
structure of this hadronic state makes further studies of its properties very
promising for understanding the strong dynamics of heavy mesons. At present the
$X(3872)$ resonance is observed experimentally only in the decays of $B$ mesons
$B \to X \, K$\,\cite{belle0,babar} and in inclusive production in proton -
antiproton collisions at the Tevatron\cite{cdf,d0}. Both these types of
processes are quite rare and also present significant challenges for precision
measurements of the parameters of the discussed resonance. In
particular\cite{pdg}, neither the total width of $X(3872)$ is yet resolved (the
current limit is $\Gamma_X < 2.3\,$MeV), nor its mass is known with a precision
sufficient to determine the binding energy of the molecular component (the
current average value $M_X = 3871.2 \pm 0.5\,$MeV coincides within the errors
with $M(D^0)+M(D^{*0})=3871.2 \pm 0.8\,$MeV). In this paper we consider the
process $e^+e^- \to \gamma X(3872)$ at the c.m. energy within few MeV of the
$D^{*0} {\bar D}^{*0}$ threshold, where the kinematical simplicity of the
process would hopefully allow more detailed studies of $X(3872)$. We estimate
that the cross section $\sigma[e^+e^- \to \gamma X(3872)]$ is likely to be at
least about $10^{-3}$ of the cross section for the production of $D^* {\bar
D}^*$ meson pairs, i.e. in the range of about 1\,pb, which makes realistic a
study of the discussed here process in a dedicated experiment. Moreover, the
energy dependence of the cross section is sensitive to the binding energy of the
molecular component. Thus a study of this dependence can provide a better
accuracy of determining the mass of $X(3872)$ relative to the $D^0 {\bar
D}^{*0}$ threshold than a direct mass measurement.

In order to estimate the cross section of the discussed process we calculate the
absorptive part of the production amplitude due to the process $e^+e^- \to
D^{*0} {\bar D}^{*0} \to \gamma \, X(3872)$ with on-shell $D^*$ mesons, as given
by the unitarity relation. We find that this contribution to the amplitude is
rapidly changing with the c.m. energy of the process. The contribution of other
intermediate states to the amplitude, which potentially could destructively
interfere with the calculated amplitude is a slowly varying function of energy,
so that such destructive interference cannot occur at all energies in the
considered range. Thus the value of the cross section can be estimated as being
at least that given by the calculated part of the amplitude.

Proceeding to the calculation of the absorptive part of the amplitude we write
the Fock decomposition of the wave function of $X(3872)$ in the form
\beq
\Psi_X= a_0 \, {D^0 {\bar D}^{*0} + D^{*0} {\bar D}^0 \over \sqrt{2}} + \sum_i
a_i \psi_i~,
\label{fock}
\eeq
where the molecular component is explicitly separated, and the sum runs over
`other' states. The crucial difference between the molecular and `other'
components of $X(3872)$ is that due to a very small binding energy $w$ the
neutral $D$ and $D^*$ mesons mostly move at long distances, beyond the range of
the strong interaction, while the `other' states are localized at shorter
distances typical of the strong interaction. At long distances the coordinate
wave function of the meson pair is that of a free $S$-wave motion and is
proportional to $\exp(-\kappa r)/r$, where $\kappa=\sqrt{2 \, m_r \, w} \approx
44 \, {\rm MeV} \, \sqrt{w ( {\rm MeV})}$ is determined by the  reduced mass
$m_r \approx 966\,$MeV in the $D^0 {\bar D}^{*0}$ system and the binding energy
$w$. With the coordinate wave function of this component normalized to one, the
coefficient $a_0$ in the expansion (\ref{fock}) determines the statistical
weight $|a_0|^2$ of the molecular state $(D^0 {\bar D}^{*0} + D^{*0} {\bar D}^0)
/ \sqrt{2}$ in the wave function of $X(3872)$. The notion of the resonance $X$
being mostly a molecular system corresponds to this statistical weight factor of
order one. The numerical value of $|a_0|^2$ is presently unknown, in a model
calculation\cite{swanson} this weight factor is estimated as 0.7 - 0.8 at $w =
1\,$MeV.

In what follows we calculate the contribution of the `peripheral'  $D^0 {\bar
D}^{*0} + D^{*0} {\bar D}^0$ component of the $X(3872)$ resonance to the
absorptive part of the amplitude of the process $e^+e^- \to  D^{*0} {\bar
D}^{*0} \to \gamma \, X(3872)$, where the latter transition proceeds due to the
underlying radiative decay $D^{*0} \to D^0 \gamma$  (${\bar D}^{*0} \to {\bar
D}^0 \gamma$), in analogy with the previously discussed\cite{mv,mv2} decays of
the $X$ resonance, $X \to D{\bar D}\gamma$. The amplitude of the radiative decay
of the vector meson has the form
\beq
A(D^{*0} \to D^0 \gamma)=\mu \, \epsilon_{ijk} \,\varepsilon_i^*  \, k_j
\, a_k~,
\label{ad0}
\eeq
with ${\vec a}$ and ${\vec \varepsilon}$ being the polarization amplitudes of
the vector meson and the photon, and ${\vec k}$ being the photon
momentum.\footnote{The non-relativistic normalization for the heavy meson states
is used throughout this paper.} The vector meson decay rate $\Gamma_0 \equiv
\Gamma(D^{*} \to D \gamma)$ is then related to the transition magnetic parameter
$\mu$ as
\beq
\Gamma_0 = {|\mu|^2 \, \omega_0^3
\over 3 \pi}~,
\label{gd0}
\eeq
where $\omega_0 \approx 137\,$MeV is the photon energy in the decay. Numerically
the decay rate can be estimated from the data\cite{pdg} as $\Gamma_0=26 \pm
6\,$KeV. The amplitude of the transition $D^{*0} {\bar D}^{*0} \to \gamma \,
X(3872)$ due to the `peripheral' component of the $X$ can then be written in
terms of the momentum-space wave function $\phi({\vec q})$ of this component:
\beq
A(D^{*0} {\bar D}^{*0} \to \gamma \, X)= {\mu \, a_0 \over \sqrt{2}} \,
\epsilon_{ijk} \,\varepsilon_i^*  \, k_j
\, \left [  a_k~ \, ({\vec b} \cdot {\vec \chi}^*) \, \phi\left( {\vec p} -
{{\vec k} \over 2} \right ) - b_k~ \, ({\vec a} \cdot {\vec \chi}^*) \,
\phi\left( {\vec p} + {{\vec k} \over 2} \right ) \right ]~,
\label{amp1}
\eeq
where, in addition to the notation conventions in Eq.(\ref{ad0}), ${\vec b}$
stands for the polarization amplitude of the initial ${\bar D}^*$ meson and
${\vec \chi}$ is the polarization amplitude of the produced $X$ resonance. The
relative minus sign between the two terms originating from the amplitudes of the
processes $D^{*0} \to D^0 \gamma$ and ${\bar D}^{*0} \to {\bar D}^0 \gamma$ is
due to the opposite C parity of the $X$ ($J^{PC}=1^{++}$) and of the
photon\cite{mv,mv2}.

In the present calculation we describe the peripheral component by a wave
function of a free-motion with an ultraviolet regularization at large
momenta\cite{suzuki,mv2}:
\beq
\phi({\vec q})= \sqrt{8 \pi \kappa} \, c \, \left ( {1 \over {\vec q}^{\,2} +
\kappa^2} - {1 \over {\vec q}^{\,2} + \Lambda^2} \right )~,
\label{phir}
\eeq
where the normalization constant $c$ is given by
\beq
c={\sqrt{\Lambda \, (\Lambda+\kappa)} \over \Lambda - \kappa}~,
\label{cnorm}
\eeq
and the regularization parameter $\Lambda$ is determined by the inverse size of
the strong interaction region in the $X$ resonance. The obvious reason for
introducing the cutoff $\Lambda$ is that the free-motion description of the
meson pair inside $X$ is applicable only at distances beyond the range of the
strong interaction and such description generally fails at short distances,
where the mesons overlap and the meson pair strongly mixes with the `other'
states in the expansion (\ref{fock}). Thus introducing the parameter $\Lambda$
is a way of explicitly separating the `peripheral' part from the `core'.

The other ingredient in the calculation of the amplitude of the process $e^+e^-
\to  D^{*0} {\bar D}^{*0} \to \gamma \, X(3872)$ is the production amplitude for
the $D^* {\bar D}^*$ pair in $e^+e^-$ annihilation. At energy $E=2M(D^{*0})+W$
near the threshold this amplitude can be generically written in the form
\beq
A(e^+e^- \to D^{*0} {\bar D}^{*0})= A_0 \, ({\vec j} \cdot {\vec p}) \, ({\vec
a} \cdot {\vec b})^* + {3 \over 2 \, \sqrt{5}} \, A_2 \, j_i \, p_k \, \left [
a_i \, b_k + a_k \, b_i - {2 \over 3} \, \delta_{ik} \, ({\vec a} \cdot {\vec
b}) \right ]^*~,
\label{aee}
\eeq
where ${\vec j}= ({\bar e} {\vec \gamma} e)$ stands for the current of the
incoming electron and positron, ${\vec p}$ is the momentum of one of the mesons
($D^{*0}$ for definiteness) in the c.m. frame, and $A_0$ and $A_2$ are the
factors corresponding to production of the vector meson pair in the states with
respectively the total spin $S=0$ and $S=2$. It can be also noted that the
amplitude in Eq.(\ref{aee}) describes the production of mesons in the $P$ wave.
Another kinematically possible amplitude, the $F$-wave, should be small near the
threshold, i.e. at a small $W$. Both $A_0$ and $A_2$ are generally functions of
the excitation energy $W$. Furthermore, their dependence on the energy near the
threshold is known to be nontrivial due to the $\psi(4040)$ resonance\cite{pdg},
with possible further complications in the immediate vicinity of the
threshold\cite{poling,dv}. Neither the relative magnitude nor the relative phase
of the amplitudes $A_0$ and $A_2$ is presently known, but both of these can be
measured from angular correlations\cite{mv3}. These amplitudes determine the
total cross section for production of $D^{*0} {\bar D}^{*0}$ in $e^+e^-$
annihilation:
\beq
\sigma(e^+e^- \to D^{*0} {\bar D}^{*0}) = \int |A(e^+e^- \to D^{*0} {\bar
D}^{*0})|^2 \, 2\pi \, \delta \left ( W-{p^2 \over m}\right ) \, {d^3 p \over (2
\pi)^3} = C \, {m \, p^3 \over 2 \pi} \, \left ( |A_0|^2 + |A_2|^2 \right )~,
\label{sdd}
\eeq
where $m = M(D^{*0})$, $p=|{\vec p}|$, and $C$ is an overall constant related to
the average value of the current $|{\vec j}|^2$. The specific value of the
latter constant will not be essential in further calculation, since it cancels
in the ratio of the cross sections. It should be pointed out that the
nonrelativistic expression for the phase space is used in Eq.(\ref{sdd})
corresponding to the nonrelativistic normalization of the states of the heavy
mesons.

The discussed here absorptive part of the amplitude of the process $e^+e^- \to
\gamma X(3872)$ due to the $D^{*0} {\bar D}^{*0}$ intermediate state is found
from the unitarity relation and the amplitudes (\ref{amp1}) and (\ref{aee}) in
the standard way:
\begin{eqnarray}
\label{abs}
&&A_{\rm Abs}(e^+e^- \to \gamma X) =  \\ \nonumber
&& {1 \over 2} \int \, \sum_{pol} \, A(e^+e^- \to D^{*0} {\bar D}^{*0}) \,
A(D^{*0} {\bar D}^{*0} \to \gamma \, X) \, \, 2\pi \, \delta \left ( W-{p^2
\over m}\right ) \, {d^3 p \over (2 \pi)^3} = \\ \nonumber
&&{\mu \, a_0 \, p \, m \over 2 \, \omega^2} \, \sqrt{\kappa \over \pi} \, F \,
\epsilon_{ijk} \,\varepsilon_i^*  \, k_j \left [ \chi_k^* \, ({\vec j} \cdot
{\vec k}) \, \left (A_0 - {A_2 \over \sqrt{5}} \right ) + j_k \, ({\vec k} \cdot
{\vec \chi}^*) \, {3 \over 2 \, \sqrt{5}} \, A_2 \right ]~.
\end{eqnarray}
In the latter expression $\omega=|{\vec k}|$ is the energy of the photon, the
sum goes over the polarizations of the vector mesons in the intermediate state,
and $F$ stands for the dimensionless form factor:
\beq
F= {1 \over \sqrt{8 \pi \kappa} } \, \int_{-1}^1 \, ({\vec p} \cdot {\vec k}) \,
\phi \left( {\vec p} - {{\vec k} \over 2} \right ) \, d \cos \theta~,
\label{f0}
\eeq
where $\theta$ is the angle between the vectors ${\vec p}$ and ${\vec k}$. Using
the expression (\ref{phir}) for the `peripheral' wave function, one readily
finds the form factor as
\beq
F= { c \over p \, \omega} \left [ \left ( p^2 +{\omega^2 \over 4} + \kappa^2
\right) \, \ln { (p+\omega/2)^2+ \kappa^2 \over (p- \omega/2)^2+ \kappa^2} -
  \left ( p^2 +{\omega^2 \over 4} + \Lambda^2 \right) \, \ln { (p+\omega/2)^2+
\Lambda^2 \over (p- \omega/2)^2+ \Lambda^2} \right ]
\label{f1}
\eeq
with the normalization coefficient $c$ given by Eq.(\ref{cnorm}).

The absorptive part of the amplitude in Eq.(\ref{abs}) corresponds to the cross
section
\begin{eqnarray}
&&\sigma_{\rm Abs} (e^+e^- \to  D^{*0} {\bar D}^{*0} \to \gamma  X) = \int
|A_{\rm Abs}(e^+e^- \to \gamma X)|^2 \, 2 \pi \, \delta(\omega - |{\vec k}|) \,
{d^3 k \over (2 \pi)^3 \, 2 \omega} =  \nonumber \\
&&C {|\mu|^2 \, |a_0|^2 \, p^2 \, m^2 \, \omega \, \kappa \, F^2 \over 12 \pi^2}
\, \left ( \left |  A_0- {A_2 \over \sqrt{5}} \right |^2 + {9 \over 20} \left |
A_2 \right |^2  \right )~,
\label{sabs}
\end{eqnarray}
where the overall constant $C$  in the latter expression is the same as in
Eq.(\ref{sdd}). Thus using also Eq.(\ref{gd0}) one finds the formula for the
ratio of the cross sections:
\beq
{\sigma_{\rm Abs} (e^+e^- \to  D^{*0} {\bar D}^{*0} \to \gamma  X) \over \sigma
(e^+e^- \to  D^{*0} {\bar D}^{*0})} = |a_0|^2 \, {\Gamma_0 \, m \, \omega \,
\kappa \over 2 \omega_0^3 \, p} \, F^2 \, {  |  A_0- {A_2/ \sqrt{5}}  |^2 +
(9/20) \,  | A_2  |^2  \over |A_0|^2 + |A_2|^2 }~.
\label{sabsr}
\eeq
The so-defined cross section $\sigma_{\rm Abs}$ is (most likely) not the actual
value of the cross section, since the amplitude of the process $e^+e^- \to
\gamma X$ can receive contribution from other mechanisms. Nevertheless
it is instructive to examine the numerical value and the behavior with energy of
this quantity as given by Eq.(\ref{sabsr}). The dependence on the c.m. energy of
the factor $(\kappa/p) \, F^2$ is shown in Fig.1 for two representative values
of the `molecular' binding energy in $X(3872)$, $w=1\,$MeV ($\kappa \approx
44\,$MeV) and $w=0.3\,$MeV ($\kappa \approx 24\,$MeV). This factor peaks at the
energy where $p \approx \omega/2 \approx 70\,$MeV. The appearance of this peak
is easily understood qualitatively: at ${\vec p} \approx {\vec k}/2$ the $D^0$
meson emerging from the emission of the photon in $D^{*0} \to D^0 \gamma$  moves
slowly relative to the ${\bar D}^{*0}$ and forms a loosely bound
state.\footnote{The same situation arises at ${\vec p} \approx -{\vec k}/2$ for
the ${\bar D}^0$ meson emerging from ${\bar D}^{*0} \to {\bar D}^0 \gamma$.} The
width of the peak is clearly determined by the parameter $\kappa$.

\begin{figure}[ht]
\begin{center}
 \leavevmode
    \epsfxsize=10cm
    \epsfbox{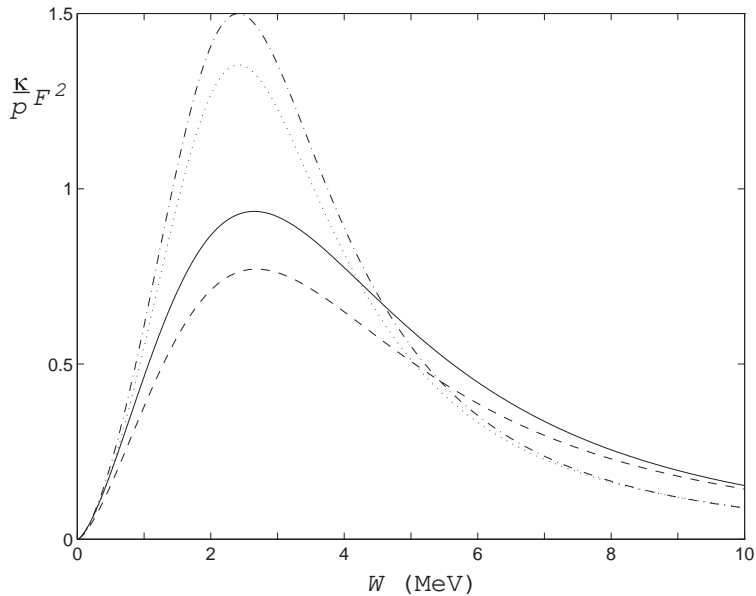}
    \caption{The factor $\kappa \, F^2/p$  vs. the excitation energy $W$ above
the $D^{*0} {\bar D}^{*0}$ threshold  at
representative values of the binding energy $w$ in $X(3872)$ and the ultraviolet
cutoff parameter $\Lambda$: $w=1\,$MeV, $\Lambda=200\,$MeV (solid), $w=1\,$MeV,
$\Lambda=300\,$MeV (dashed), $w=0.3\,$MeV, $\Lambda=200\,$MeV (dashdot), and
$w=0.3\,$MeV, $\Lambda=300\,$MeV (dotted).}
\end{center}
\end{figure}

As is seen from the plots of Fig.1 the numerical value of the factor $(\kappa/p)
\, F^2$ near its peak is of order one. Another factor in Eq.(\ref{sabsr}),
$\Gamma_0 \, m \, \omega /(2  \omega_0^3) \approx \Gamma_0 \, m /(2  \omega_0^2)
\approx 1.5 \times 10^{-3}$, sets the overall scale of the discussed cross
section. The factor in Eq.(\ref{sabsr}) depending on the presently unknown ratio
of the (generally complex) amplitudes $A_0/A_2$, takes values between 0.34 (at
$A_0/A_2 \approx 0.68$) and 1.31 (at $A_0/A_2 \approx 1.47$), and can thus be
considered as being of order one. Finally, the statistical weight factor
$|a_0|^2$, as discussed, is likely to be large fraction of one. Summarizing
these numerical estimates, the value of the ratio in Eq.(\ref{sabsr}) at the
peak can be estimated as being of order $10^{-3}$, although the uncertainty is
presently large.

In absolute terms, the measured\cite{poling} cross section $\sigma(e^+e^- \to
D^{*0} {\bar D}^{*0})$ at $E = 4015\,$MeV, i.e. at the energy above the $D^{*0}
{\bar D}^{*0}$ threshold $W \approx 1.6\,$MeV is about 0.15\,nb. This cross
section grows from the threshold as $p^3$. With this factor taken into account
the peak of the quantity $\sigma_{\rm Abs} (e^+e^-  \to \gamma  X)$ shifts to a
slightly higher value of $p$, $p \approx 100\,$MeV, corresponding  to $W \approx
5\,$MeV, where according to Eq.(\ref{sabs}) and the presented estimates, it
should be numerically of the order of 1\,pb.

The considered mechanism of the process $e^+e^- \to \gamma X$ describes a `soft'
production of its peripheral $D^0 {\bar D}^{*0} + D^{*0} {\bar D}^0$ component
in radiative transitions from slow $D^{*0} {\bar D}^{*0}$ pairs. Generally, one
can also expect a presence of states with charged mesons, $D^+ D^{*-}+D^-
D^{*+}$, within the $X$ resonance. Therefore a contribution of the mechanism
$e^+e^- \to D^{*+} D^{*-} \to \gamma X$ to the considered here process merits
discussion.  However, this contribution in fact should be very small for at
least three reasons\cite{mv2}: The mass of the pair of charged mesons is by
$\delta \approx 8\,$MeV heavier than that of the neutral ones. For this reason
the charged mesons in this component are separated by shorter distances: the
corresponding parameter $\kappa$ is approximately $\sqrt{2 m_r \delta} \approx
125\,$MeV, and as a result\cite{mv2} the statistical weight of such state should
be significantly smaller than for the pair of neutral mesons. Furthermore, the
C-conjugate processes $D^{*+} \to D^+ \gamma$ and $D^{*-} \to D^- \gamma$
destructively interfere in $D^{*+} D^{*-} \to \gamma X$ (cf. the minus sign
between the two terms in Eq.(\ref{amp1})). The (negative) interference is
enhanced for the more closely separated charged mesons in $X(3872)$. Finally,
the transition magnetic coupling $\mu$ is noticeably weaker for the charged
mesons than for the neutral ones: $\Gamma(D^{*+} \to D^+ \gamma) = 1.5 \pm
0.5\,$KeV.

Other intermediate states with charmed meson pairs, i.e. $D {\bar D}$ and $D
{\bar D}^*$ (${\bar D} D^*$),   can potentially contribute to the discussed
process $e^+e^- \to \gamma X$. Indeed, the suitable final state arises in the
chain $e^+e^- \to D {\bar D} \to \gamma \, (D {\bar D}^* +D^* {\bar D})$ through
the radiative transition $D \to \gamma D^*$ ( ${\bar D} \to \gamma {\bar D}^*$)
as well as in the chain $e^+e^- \to D {\bar D}^*+{\bar D} D^* \to \gamma \, (D
{\bar D}^* +D^* {\bar D})$ through an elastic emission of a photon by $D^*$
(${\bar D}^*$). However, one can readily see that in either of these processes
the charmed meson emerging after the emission of the photon is very far off the
mass shell in the scale of $\kappa$. Thus neither of these processes can proceed
due to the long-distance peripheral component of the $X(3872)$ resonance, but
rather is determined by the short-distance dynamics of the `core' of $X$. For
this reason these contributions, as well as other possible mechanisms related to
the `core' dynamics, should be smooth functions of the c.m. energy on the scale
of few MeV around the $D^{*0} {\bar D}^{*0}$ threshold, where the amplitude
given by Eq.(\ref{abs}) experiences a significant variation. Therefore even
under the most conservative (and quite unlikely) assumption that these
mechanisms cancel the contribution of the latter amplitude near its maximum,
such cancellation cannot take place at all energies in the considered range.
Thus the cross section of the process $e^+e^- \to \gamma X$ at an energy within
few MeV of the $D^{*0} {\bar D}^{*0}$ threshold has to be at least as large as
the above estimates for $\sigma_{\rm Abs}$ near its maximum i.e. of the order of
1\,pb. The latter is a conservative estimate, since we cannot exclude that the
contribution of those `other' mechanisms exceeds the calculated amplitude and
that the actual cross section is larger than $\sigma_{\rm Abs}$.

The work of MBV is supported, in part, by the DOE grant DE-FG02-94ER40823.


\begin{thebibliography}{99}
\bibitem{pdg}
  W.~M.~Yao {\it et al.}  [Particle Data Group],
  J.\ Phys.\ G {\bf 33}, 1 (2006).
\bibitem{belle}
K. Abe {\it et.al.}, [Belle Coll], Report BELLE-CONF-0540, May 2005,\
[hep-ex/0505037].
\bibitem{cp}
F.E. Close and P.R. Page, Phys.Lett.B {\bf 578}, 119 (2004).
\bibitem{ps}
S. Pakvasa and M. Suzuki, Phys.Lett.B. {\bf 579}, 67 (2004).
\bibitem{mv}
M.B. Voloshin, Phys.Lett.B. {\bf 579}, 316 (2004).
\bibitem{nat2}
N.A. T\"ornqvist, Phys.Lett.B {\bf 590}, 209 (2004);\ [hep-ph/0402237
] and \ [hep-ph/0308277].
\bibitem{ov}
M.B. Voloshin and L.B. Okun, JETP Lett. {\bf 23}, 333 (1976).
\bibitem{belle0}
  S.~K.~Choi {\it et al.}  [Belle Collaboration],
  Phys.\ Rev.\ Lett.\  {\bf 91}, 262001 (2003)
  [arXiv:hep-ex/0309032].
\bibitem{babar}
  B.~Aubert {\it et al.}  [BABAR Collaboration],
  Phys.\ Rev.\ D {\bf 73}, 011101 (2006)
  [arXiv:hep-ex/0507090].
\bibitem{cdf}
  D.~Acosta {\it et al.}  [CDF II Collaboration],
  Phys.\ Rev.\ Lett.\  {\bf 93}, 072001 (2004)
  [arXiv:hep-ex/0312021].
\bibitem{d0}
  V.~M.~Abazov {\it et al.}  [D0 Collaboration],
  Phys.\ Rev.\ Lett.\  {\bf 93}, 162002 (2004)
  [arXiv:hep-ex/0405004].
\bibitem{swanson}
  E.~S.~Swanson,
  Phys.\ Lett.\ B {\bf 588}, 189 (2004)
  [arXiv:hep-ph/0311229].
\bibitem{mv2}
  M.~B.~Voloshin,
  Int.\ J.\ Mod.\ Phys.\ A {\bf 21}, 1239 (2006)
  [arXiv:hep-ph/0509192].
\bibitem{suzuki}
  M.~Suzuki,
  Phys.\ Rev.\ D {\bf 72}, 114013 (2005)
  [arXiv:hep-ph/0508258].
\bibitem{poling}
  R.~Poling,
  eConf {\bf C060409}, 005 (2006)
  [arXiv:hep-ex/0606016].

\bibitem{dv}
  S.~Dubynskiy and M.~B.~Voloshin,
  arXiv:hep-ph/0608179.
\bibitem{mv3}
  M.~B.~Voloshin,
  eConf {\bf C060409}, 014 (2006)
  [arXiv:hep-ph/0605063].



\end{thebibliography}
\end{document}